\documentclass[12pt,a4paper]{article}
\usepackage[dvips]{graphics}
\begin{document}
  \title{Diploid versus Haploid Organisms}
  \author{Armando Ticona and Paulo Murilo C. de Oliveira\\
Instituto de F\'{\i}sica, Universidade Federal Fluminense\\ Av. Litor\^{a}nea s/n, Boa Viagem, Niter\'{o}i 24210-340, RJ, Brazil\\
e-mail: aticona@if.uff.br, pmco@if.uff.br}
  \maketitle

\begin{abstract}
Using a bit string model, we show that asexual reproduction for diploids is more efficient than for haploids: it improves genetic material producing new individuals with less deleterious mutations. We also see that in a system where competition is present, diploids dominate, even though we consider some dominant loci.
\end{abstract}


\section{Introduction}\label{sec:int}

In a system where population is kept constant, competiton just allows the best individuals to survive, i.e. individuals which have less deleterious mutations and are better adapted to the current enviroment have a higher probability to survive. Haploid gametes reproduce by mitosis, mutations being the only source of variability, thus they cannot improve too much when reproducing. On the other hand, diploids could reproduce with crossings between two equivalent copies of genetic material, by meiotic parthenogenesis. This is supposed to be a more efficient way to improve the ability to adapt the population to the environment. The purpose of this work is to verify this hypothesis within a simple bit string model \cite{ref1} of population dynamics.

The hypothesis of Jan, Stauffer and Moseley (JSM)\cite{ref2} proposes that haploid populations convert first into asexual diploid and then diploids begin to reproduce in a sexual way. This complete process has been studied in references \cite{ref3,ref4}, although they used a completely different model, with conversions (haploid $\rightarrow$ diploid and asexual $\rightarrow$ sexual) happening only for individuals with many deleterious mutations. In this case, sexual diploids finally dominate the system with a small fraction of asexual diploids and haploids.

In this work we show that in a system under competition, with an asexual population completely haploid at the beginning, and where with a small probability two haploids can fuse to create a diploid, evolution leads to a population dominated by diploids at the end. We also show that even the introduction of some dominant loci into the genetic material of diploids does not change too much this final result.

\section{Models for the haploid-diploid cycle}\label{sec:mod}

We represent each haploid by a bit string of 1024 ``0"'s and ``1"'s, with a probability to survive given by: $p_i = x^{N_i+1}$, where $N_i$ is the number of ``1"'s in the bit string representing individual $i$. Every time step each haploid reproduces by duplicating its bit string and some $M_H$ random mutations are performed in this new individual, say $M_H = 1$, by flipping $M_H$ randomly chosen bits from 0 to 1 or vice-versa.

After this breeding process applied to all individuals, in  order to keep the population almost constant, fluctuating around some number $P_o$,  say $P_o = 10000$, we determine the value of $x$ by solving equation:

\begin{equation}\label{eq1}
\sum_{N = 0}^{\infty} H(N)x^{N + 1} = P_o
\end{equation}
where $H(N)$ is the number of haploids with $N$ ``1"'s into their bit string. Already knowing the value of $x$, we test each individual $i$ of the population, to keep it alive according to the probability $x^{N_i+1}$, killing it otherwise. All this represent one time step, where all individuals reproduce first and then each one is tested to survive (including newborns).

For a population of diploids, we represent each individual by two parallel bit strings of 1024 ``0"'s and ``1"'s. At each time step, every diploid reproduce by meiotic parthenogenesis. Both strings are cut in the same random position and a new bit string is constructed by joinning one randomly chosen piece with the complementary piece of the other string. Then, we copy this new bit string in order to have two bit strings representing this new diploid. Finally, $M_D$ mutations, say $M_D = 2$, are randomly performed.

The probability to survive in this case is also given by: $p_i = x^{N_i+1}$, but now $N_i$ represents the number of coincident positions with ``1"'s in both bit strings (homozygous loci). We can also consider some dominant positions along the bit strings: in this case, just one ``1" bit at this position (locus) is enough to contribute to $N_i$.

\subsection{Populations without exchange}\label{sse:nex}
First we study how two populations (one haploid and the other diploid) develop living together under the same environment, but with no possibility of haploids becoming diploids or vice-versa. We begin a population of $P_o/2$ haploids and $P_o/2$ diploids. At each time step, both kinds of individuals reproduce following the rules already quoted, using the same $x$, and we perform the sum of equation (\ref{eq1}) over all haploids and diploids in order to keep the whole population constant.

\begin{figure}[hp]
\centering
\scalebox{0.4}{\includegraphics{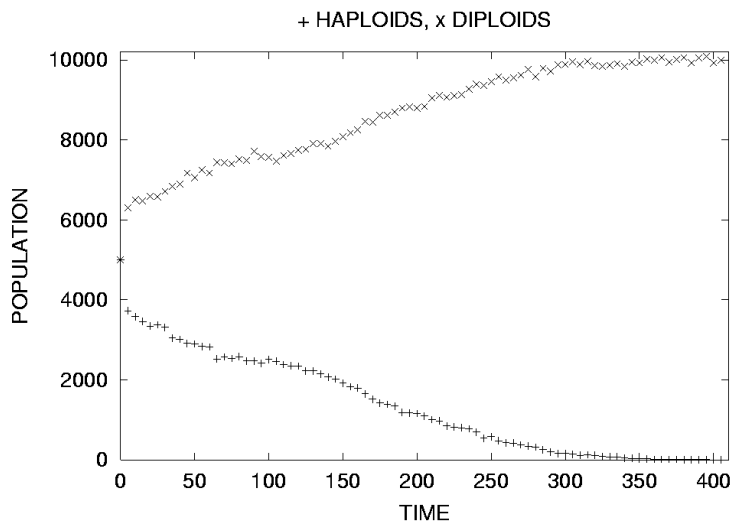}}
\caption{Haploids versus Diploids}
\label{fig1}
\end{figure}

Figure 1 shows the number of haploids and diploids evolving with time. Most of the time diploids dominate the whole system, and eventually no haploid survives. The same behaviour was observed for different runs.

This situation does not change too much if we include some dominant positions for diploids, that is, less than 50\% of dominant positions. Only for an unrealistic situation where more than 50\% of positions are dominant, we can see some instances where the haploids dominate the system.

\subsection{Exchange (diploids with only recessive positions)}\label{sse:res}

\subsubsection{No return}\label{sss:nrr}
Now we begin with a population totally haploid, but each individual has a small probability $P_{HD}$, say $P_{HD} = 0.01$, to randomly choose another individual to fuse and form a diploid, instead of normal reproduction. The fused diploid has the two bit strings of those haploids. Otherwise, this haploid reproduces in the normal way. During the evolution, diploids also reproduce according to the same rules explained before.

In this case we can see that the population becomes totally diploid, after few steps. Using different values of $P_{HD}$ we just change the number of steps needed to converge to an only-diploid population.

\subsubsection{With return}\label{sss:wrr}
Now, we add some probability $P_{DH}$, that a diploid transform itself into two haploids, each one with one of the bit strings from the diploid, with some $M_H$ mutations.

In Nature this backward transformation is supposed to be harder to happen than the first case (haploid $+$ haploid $\rightarrow$ diploid). The diploid organism needs to divide its genetic material in just two equal parts, otherwise at least the smallest part which misses genetic material will die before reproduction. 

\begin{figure}[hp]
\centering
\scalebox{0.4}{\includegraphics{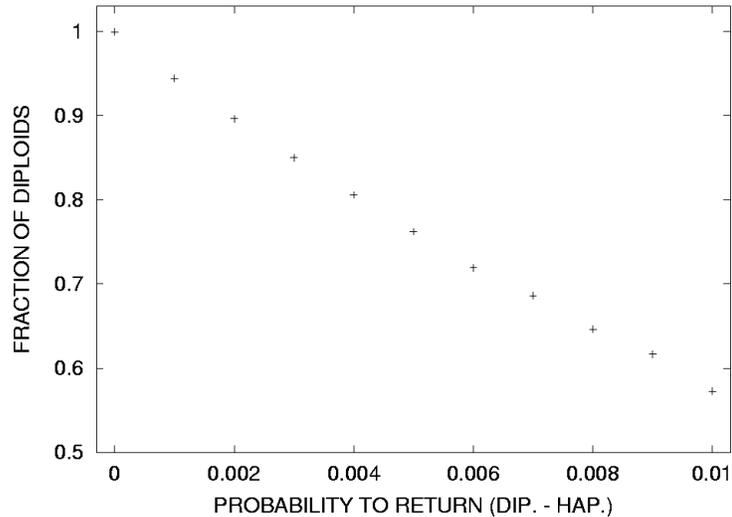}}
\caption{Fraction of diploids in population as a function of $P_{DH}$}
\label{fig2}
\end{figure}

For this case we observed that diploids still dominate the system but there is a small fraction of haploids at the end, which are products of the continuous transformation of some diploids into haploids. In figure 2 we see the fraction of diploids in the population as a function of the back probability $P_{DH}$. These results are obtained as an average over many steps, after equilibrium has been reached. Althougth some extreme values of $P_{DH}$ are taken (as $P_{DH} = P_{HD}$ corresponding to the rightmost point), we can see that diploid individuals always dominate the population.

\subsection{Exchange (diploids with recessive and dominant positions)}\label{sse:dom}

\subsubsection{No return}\label{sss:nrd}
Now, we repeat the same experiment explained before, but consider some percentage $D$ of dominant loci for the diploids. As before, some haploids become diploids at each step, with no diploid becoming haploid. As a result, again, after some steps the whole population becomes diploid with no haploids in the final equilibrium configuration. This result is independent of the value of $D$: even an extreme value of $D$ ($D = 100\%$) just delay the final situation by some steps.

\subsubsection{With return}\label{sss:wrd}
If we let diploids to return back to haploids with some probability $P_{DH}$ and we also consider some dominant positions for diploids, the results obtained before do not change qualitatively, as we can see in figure 3. The fraction of diploids at the equilibrium situation is plotted for different values of $D$, and for fixed values of $P_{HD} = 0.01$ and $P_{DH} = 0.005$, taken as an example.

\section{Discussion}\label{sec:dis}
In order to study the most efficient kind of behaviour (haploid versus diploid), we considered systems where competiton allows to survive only a pre-determi\-ned number of individuals. Of course, these selected individuals will be the best adapted to the environment, presenting a small number of deleterious mutations.

Diploid reproduction is more efficient to improve new generations than haploid reproduction. The final population is dominated by diploids, and haploids are present only if we let some diploids to become haploids.

The presence of some dominant loci in diploids does not change the results. In the limiting case $P_{DH} = 0$ (no diploid returns back to haploid condition), dominance just modifies a little bit the equilibration time, but eventually we have no haploids anyway.

\begin{figure}[hp]
\centering
\scalebox{0.4}{\includegraphics{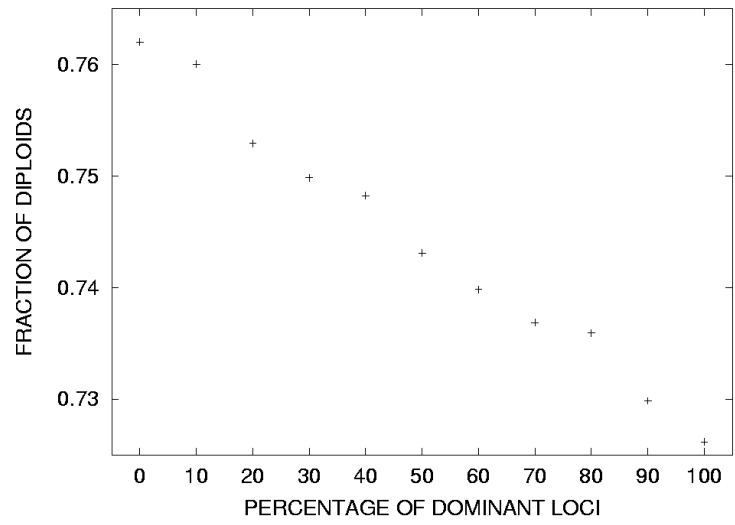}}
\caption{Fraction of diploids in population as a function of $D$}
\label{fig3}
\end{figure}

\paragraph{Acknowledgements}
We thank Dietrich Stauffer for his comments about this work.

\end{document}